# Alteration in contact line dynamics through competition between thermocapillary and electrothermal effects

Golak Kunti, Anandaroop Bhattacharya, Suman Chakraborty,[a]

Department of Mechanical Engineering, Indian Institute of Technology Kharagpur, Kharagpur, West Bengal - 721302, India

[a]*E-mail address* of corresponding author*:* suman@mech.iitkgp.ernet.in

**ABSTRACT**

The contact line dynamics of a thermal field assisted flow configuration of two immiscible fluids is considered in a narrow thermofluidic pathway. The surfaces of the channel are wetted with predesigned wettabilities and interdigitated electrodes are mounted on the substrates to generate non-uniform electric field. In this study, the interplay of thermocapillary and electrothermal forces on interfacial dynamics are studied in detail. The former is caused by surface tension gradients while the latter results from gradients in permittivity and electrical conductivity. Our investigations reveal that the relative strength of interfacial forces and electrothermal forces and their interactions can be effectively used to control the capillary filling time as well as the flow dynamics. For same strength of thermocapillary and electrothermal forces (characterized by individual dimensionless numbers) electrothermal forces dominate over thermocapillary forces. However, interfacial forces at some wettabilities dominate over electrothermal forces and set the fluid motion toward the entry of the channel.

## I. Introduction

The dynamics of contact line of fluid-fluid-solid interface and its intricate motion over wetted substrates possess several industrial applications and occur in a numerous number of naturally occurring phenomena such as drops of rain flowing over window panes and leaves, flow of magma, tear films and so on. Manipulation of biochemical reactions, biochemical analysis, chemical synthesis, clinical diagnostics and drug delivery are common processes and are encountered in microfluidic devices [1–5]. Inherent interesting physics of two-phase flow has received a deep attention in those miniaturized devices [6,7]. Typically chemical technologies, polymer processing, oil recovery, powder wetting, textile manufacturing, microscale biomedical processes and biochemical analysis, manufacturing of photographic films are the areas of interest where motion of contact line of two immiscible fluids appears frequently [8–13]. In modern days, technological development and reliability in manufacturing smart devices promote careful attention on fundamental insights and attractive features of moving contact lines in research community from various disciplines.

A few decades ago, singularities in non-integrable stress created a lot of problems to formalize the contact line motion over solid surfaces [14,15]. The imposed no slip boundary condition over the substrates introduced stress singularities at the fluid-fluid-solid three phase contact line [16]. Many researchers proposed several models to evade this problem by addressing surface tension relaxation [10], considering slip [17,18], following adsorption model [19], diffusion interface formulation [20–23], to name a few. Of these, phase-field based diffusion interface model gains several advantages over other existing methodologies because of its relevance in describing contact line theory. In such application, this method inherently introduces slip on the three-phase contact line and removes the stress singularity problem [24], A number of various models of phase-field formalism such as surface energy formulation [23,25], geometric formulation [26], etc. do not follow direct interface tracking [27,28].

Immiscible fluid flow and dynamics of contact line over chemically wetted substrates were thoroughly investigated in the open literature based on mechanical-driven flow actuation, such as externally applied pressure gradient [21,23,24,29,30]. Mechanical actuators often deal with several problems, such as acoustic noise, frictional losses, reliability, and portability issues owing to moving elements etc. and are vulnerable for applications in various miniaturized devices. In many lab-on-a-chip and micro total analysis systems, temperature gradient, concentration gradient etc. can be used as a source of energy which generates fluid flow [31–35]. The externally applied thermal field generate local variations in thermophysical and electrical properties, which, in turn, cause voluminous body forces in the bulk of the fluid. Electrothermal flow is generated when an AC electric field is applied in an aqueous solution in presence of thermal field [36–39]. The inhomogeneities in electrical conductivity and permittivity arise owing to the temperature gradient in the bulk solution. The gradient in conductivity causes volumetric free charges and Coulomb forces while permittivity gradient generates dielectric forces [40–42]. Application of low voltage and a wide range of operating parameters, such as frequency, electrical conductivity allow electrothermal mechanism as more advantageous over DC electroosmosis and AC electroosmosis processes [43–45]. In recent years, chip-scale integrability, adequate portability, reliability of AC electrothermal (ACET) actuation have arrested special interest in research communities under the AC electrokinetics processes. Adopting electrothermal mechanism fluid pumping [31,46], mixing of fluids [47,48], and particle sorting [49] can be efficiently controlled in microfluidic devices.

Over recent years, investigations were conducted on thermal field-driven single phase fluid flow over wetted surfaces. However, underlying flow physics may alter non-trivially if the flow field comprises a binary system of two immiscible fluids, specifically, in presence of electrothermal forces arisen from combined consequences of electric and thermal fields. At one side, gradients in electrical conductivity and permittivity cause electrothermal forces. On the other hand, gradient in surface tension generates thermocapillary forces in the same domain. Therefore, depending on the direction of the gradient of the temperature and electric field strength, and polarity of the free charges the thermocapillary forces and electrothermal forces may alter the interfacial dynamics of the two phase system non-trivially. To fill up this knowledge gap in the open literature, we consider interfacial dynamics of a binary system with chemically wetted surface. The motion of the binary fluids is actuated using an externally applied thermal field, where electrothermal forces and thermocapillary forces control the flow dynamics. For a systematic interrogation of distinctive features of the contact line dynamics altered with thermocapillary and electrothermal forces, we define two force numbers, namely thermocapillary force number and ACET force number. The net force in the bulk fluid depends on wetting characteristics of the surface. We thoroughly explore the interplay between thermocapillary forces, interfacial forces and electrothermal forces.

**II. Physical system and mathematical model**

**A. Problem geometry**

In the present analysis, we consider interfacial motion of a two phase fluids (fluid A and fluid B) through a narrow channel shown in Fig. 1. The walls of the channel are designed with a predefined wettability which is manifested by specifying static contact angle $\theta_s$. Length and height of the channel are $L$ and $H$, respectively. The x-coordinate and y-coordinate run along the channel length and height, respectively. The width along the other direction is sufficiently large, so that flow configuration allows two-dimensional analysis. Due to symmetric nature of the flow actuation with respect to the mid-horizontal line, we consider only upper half of the channel where origin of the coordinate system lies at the left centre of the channel. Fluids A and B reside on the left side and right side of the interface, respectively. To set the fluids in motion externally thermal and electric fields are imposed into the solution domain. Five electrode pairs on each wall are embedded. The length of the narrow and wider electrodes are $d_1$ and $d_2$ whereas $g$ is the gap between electrodes. The distance between two electrode pairs is $2s$. Application of temperature gradient and potential distribution in the fluid system generates thermocapillary and electrothermal forces simultaneously which drive the fluids along the channel. Further, we consider initial axial position of the interface is at $L_1$. In the subsequent paragraphs, we will describe the detailed theoretical modeling of the transport system.

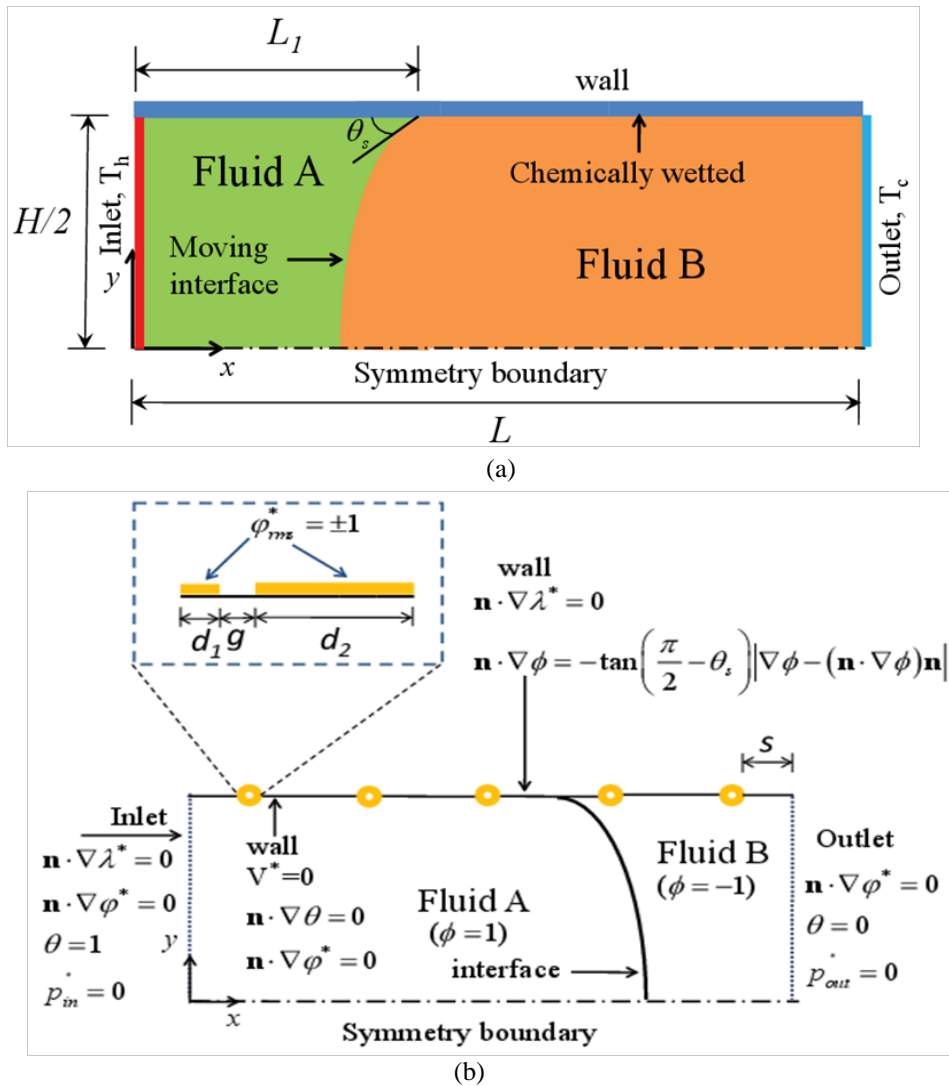

(a)

(b)

FIG. 1. (a) Physical system of the solution domain showing physical dimensions. The coordinate system is shown, the origin is taken at the centre of the left side of the channel. Due to symmetry, upper-half of the full domain is presented where Fluids A and B are occupied at left side and right side, respectively. The walls are decorated with suitable wettability condition. (b) All the boundary conditions are mentioned. Electrodes are shown in the dotted box with orange color. To imposed thermal field the inlet and outlet are maintained at nondimensional temperatures of 1 and 0, respectively. Imposed temperature gradient causes variation in properties which result in bulk volume force in the solution domain in the form of thermocapillary and electrothermal forces. Electrothermal forces are the combined consequences of the electric field and inhomogeneities in electrical properties. On activation of the various forces, fluids start moving along the channel.

**B. Theoretical modeling**

**1. Phase field formulation**

To track the interfacial motion we have used phase field method. This method is very popular and extensively used in the area of multiphase flow [6,22,29]. Strong thermodynamic basis, the capability of interface tracking implicitly, the establishment of slip phenomena at the fluid-fluid-solid interface and removal of stress singularity are the most important characteristics of this method. Phase field model works on the basis of minimization of free energy and identification of the equilibrium states of the multi phase system. Different states of the system are described using an order parameter $\phi$, also known as phase field parameter. For a two phase immiscible fluids system, the phase concentration of fluids A and B are specified by order parameter $\phi = 1$ and $\phi = -1$, respectively, whereas the interface is described by $\phi \in (1, -1)$. The thermodynamics of the system is governed by the Ginzburg-Landau free energy concept whose mathematical expression is [24,28,50,51]:

$$F = \int_{\forall} \left\{ f(\phi) + \frac{1}{2}\eta\xi |\nabla\phi|^2 \right\} d\forall, \tag{1}$$

where $\forall$ is the volume of the domain. $\eta$ is the surface tension and $\xi$ is the thickness of the interface. $f(\phi)$ is the free energy density. Another term in the integral is the free energy density at the interface. According to the phase field theory, the interface is very thin but has finite thickness. This phenomenon is taken into consideration by this term. On the other hand, immiscibility of the fluids is accounted using free energy which is also expressed in the form of double well potential as [22,52]:

$$f(\phi) = \frac{\eta}{4\xi}(1-\phi^2)^2. \tag{2}$$

The minimum values of the free energy density can be evaluated at two stable phases $\phi = \pm 1$. The order parameter distribution, which is the combined consequences of the minimization of the free energy function and mass conservation of the two phases is expressed by the Cahn-Hilliard equation (CHE) as [53,54]:

$$\frac{D\phi}{Dt} = \nabla \cdot (M \nabla \lambda), \tag{3}$$

Where $M$ is the mobility parameter, also commonly known as Onsager coefficient. It is a function of critical mobility $M_c$ [50] and order parameter $\phi$ and can be written as $M = M_c(1-\gamma\phi^2)$, where parameter $\gamma$ takes into account the dynamical characteristics of the two phase system. The rate at which $\phi$ is diffused is controlled by the $M$. The stress singularity is removed with the continuous diffusion of the order parameter across the thin interface. The derivative of $F$ is known as chemical potential $\lambda$ and is expressed by:

$$\lambda = \frac{\delta F}{\delta \phi} = f'(\phi) - \eta \xi \nabla^2 \phi. \tag{4}$$

To solve the Cahn-Hilliard equation we have used following two boundary conditions [30]:

$$\mathbf{n} \cdot \nabla \lambda = 0, \tag{5}$$

$$\mathbf{n} \cdot \nabla \phi = -\tan\left(\frac{\pi}{2} - \theta_s\right)|\nabla\phi - (\mathbf{n} \cdot \nabla\phi)\mathbf{n}|, \tag{6}$$

where $\mathbf{n}$ is the unit normal directed outward from the solid surface. The first boundary condition introduces zero flux across the walls while another one characterizes the imposed wetting condition in the form of static contact angle ($\theta_s$). Through the order parameter variation interface profile near the boundary is controlled following the second boundary condition which is also acknowledged as geometric boundary condition [30].

2. **ACET and thermocapillary forces**

In the present work on contact line dynamics, we have adopted temperature dependent electrothermal and thermocapillary forces to generate the fluid motion. In this paragraph, we describe the consequences of various forces associated in the context of thermally actuated moving contact line dynamics. As discussed previously, electrothermal forces are the resultant effect of the thermal and electric fields. We get the potential distribution following the equation [53,54]:

$$\nabla \cdot (\sigma \nabla \varphi_r) = 0, \tag{7}$$

where $\sigma$ is the electrical conductivity. $\varphi$ is the potential and subscript $r$ denotes real component. The magnetic field is neglected as we consider a quasi-electrostatic field [55,56]. Once, the electric field is generated application of temperature gradient causes electrothermal forces which can be expressed as [36,57]:

$$\mathbf{F_E} = -\frac{1}{2}\left[\left(\frac{\nabla\sigma}{\sigma} - \frac{\nabla\varepsilon}{\varepsilon}\right) \cdot \mathbf{E}\frac{\varepsilon\mathbf{E}}{1+(\omega\tau)^2} + \frac{1}{2}|\mathbf{E}|^2\nabla\varepsilon\right]. \tag{8}$$

Here, $\mathbf{E}$ is the electric field strength and $\omega$ denotes the angular frequency of the electrical signal. The charge relaxation time $\tau$ is the ratio of permittivity ($\varepsilon$) to the electrical conductivity. The first term and second term in the right hand of the Eq. (8) are known as Coulomb force and dielectric force, respectively. Since the binary system deals with incompressible fluids the electrostriction force ($0.5\nabla(\rho\mathbf{E}^2(\partial\varepsilon/\partial\rho))$) is negligible compared to other electrical forces [36]. In accordance with phase field theory, all the fluid properties are function of the order parameter. Besides, electrical conductivity and permittivity are temperature dependent. Therefore, gradients $\nabla\sigma$ and $\nabla\varepsilon$ can be obtained following the functional dependency of electrical properties with temperature and order parameter as:

$$\begin{aligned}\varepsilon(T,\phi) &= \varepsilon_A(T_0)\{1+\alpha_A(T-T_0)\}\left(\frac{1+\phi}{2}\right) + \varepsilon_B(T_0)\{1+\alpha_B(T-T_0)\}\left(\frac{1-\phi}{2}\right),\\ \sigma(T,\phi) &= \sigma_A(T_0)\{1+\beta_A(T-T_0)\}\left(\frac{1+\phi}{2}\right) + \sigma_B(T_0)\{1+\beta_B(T-T_0)\}\left(\frac{1-\phi}{2}\right),\end{aligned} \tag{9}$$

Here, a linear variation in properties with temperature ($T$) is considered [36]. We denote the reference properties with subscripts A and B for fluids A and B, respectively. $\alpha$ stands for the gradient of the permittivity whereas $\beta$ denotes gradient of the electrical conductivity. For aqueous media these values are: $\alpha \approx -0.001\,\mathrm{K}^{-1}$ and $\beta \approx 0.01\,\mathrm{K}^{-1}$ [36].

Besides electrothermal forces, interfacial stress jump across the interface results surface tension forces and thermocapillary forces. The surface tension forces are the consequence of the local change in curvature of the interface whereas thermocapillary forces arise owing to the local gradient in surface tension caused by spatial variation in temperature. The combined form of these forces is [58]:

$$\mathbf{F_S} = \frac{3\sqrt{2}}{4}\xi\left[\eta_T |\nabla\phi|^2 \nabla T - \eta_T (\nabla T \cdot \nabla\phi)\nabla\phi + \frac{\eta}{\xi^2}\lambda\nabla\phi\right], \tag{10}$$

where $\eta_T$ is the gradient of the interfacial tension. In our analysis, we consider a linear variation in surface tension with temperature. Therefore, surface tension as a function of temperature can be represented as $\eta(T) = \eta_0 + \eta_T(T - T_0)$. The temperature and surface tension corresponding to subscripts 0 are the reference temperature and reference surface tension, respectively.

## 3. Energy and Cahn-Hilliard-Navier-Stokes equations

The two phase system involves an externally applied thermal filed. In addition, applied electric field generates small amount Joule heat ($\sigma|\mathbf{E}|^2$). The temperature distribution in the system can be obtained by solving the energy equation which can be written as:

$$\rho C_p \frac{DT}{Dt} = \nabla \cdot (k\nabla T) + \sigma|\mathbf{E}|^2, \tag{11}$$

Here, $\rho$ is the mass density. $C_p$ is the specific heat and $k$ is the thermal conductivity of the fluid. Except for electrical conductivity and permittivity, other properties do not depend on temperature and are constant in each phase. Also, various phase change events, like evaporation, condensation are not considered in the study owing to the feasibility of the problem excluding those effect within the operating ranges of parameters.

Various forces, namely thermocapillary force, surface tension force and electrothermal forces which contribute in the flow actuation are added in the transport equations. Accordingly, modified Cahn-Hilliard-Navier-Stokes equation and continuity equation can be expressed as [29,59,60]:

$$\nabla \cdot \mathbf{V} = 0, \tag{12}$$

$$\rho \frac{D\mathbf{V}}{Dt} = -\nabla p + \nabla \cdot \left[\mu(\nabla\mathbf{V} + \nabla\mathbf{V}^T)\right] + \mathbf{F_S} + \mathbf{F_E}, \tag{13}$$

We do mention here that present configurations are under laminar and incompressible. The parameters $p, \mathbf{V}, \mu$ are the pressure, fluid velocity, and viscosity of the fluid, respectively.

The fluid properties which are not function of temperature can be represented in the form of order parameter as: $f = 0.5 f_A (1+\phi) + 0.5 f_B (1-\phi)$, $f$ can be $\rho, k, C_p$ or $\mu$.

## 4. Nondimensional governing equations

We use nondimensional form of governing equations to simulate the results for which following dimensionless parameters are adopted to cast the dimensional equations:

$$u^* = \frac{u}{u_0}, v^* = \frac{v}{u_0}, x^* = \frac{x}{H}, y^* = \frac{y}{H}, t^* = \frac{tu_0}{H}, \theta = \frac{T - T_0}{\Delta T}, \varphi = \frac{\varphi}{\varphi_0}, \rho^* = \frac{\rho}{\rho_0}$$
$$\mu^* = \frac{\mu}{\mu_0}, C_p^* = \frac{C_p}{C_{p0}}, k^* = \frac{k}{k_0}, \varepsilon^* = \frac{\varepsilon}{\varepsilon_0}, \sigma^* = \frac{\sigma}{\sigma_0}, p^* = \frac{pH}{\mu u_0}, \lambda^* = \frac{\lambda}{\eta_0 / H}, \omega^* = \frac{\omega}{\omega_0}$$
(14)

where the reference velocity $u_0$ can be evaluated depending on the dominating characteristics of the various forces. Comparing viscous forces with thermocapillary forces $u_0$ can be obtained as $u_0 = \xi \eta_T \Delta T / \mu_0 H$, where $\nabla T$ is the temperature difference between inlet and the outlet. Typically, for $\xi \approx 10^{-8}$ m, $\eta_T \approx -10^{-5}$ Nm$^{-1}$K$^{-1}$, $\nabla T \approx 10$ K, $\mu_0 \approx 0.001$ Pa s and $H \approx 10^{-4}$ m the order of the reference velocity is $u_0 \approx 10^{-5}$ ms$^{-1}$. On the other hand, balance between electrothermal forces and viscous forces gives $u_0 = \varepsilon_0 \varphi_0^2 \Delta T \beta / \mu_0 H$. For $\varepsilon_0 \approx 10^{-9}$ C/Vm, $\varphi_0 \approx 1$ V, $\beta \approx -0.01$ K$^{-1}$, we obtain $u_0 \approx 10^{-4}$ ms$^{-1}$. The reference properties and parameters are denoted with subscript 0. Putting these dimensionless parameters into the dimensional form of the equations one can obtain the nondimensional transport equations as:

$$\lambda^* = -Cn \nabla^2 \phi + \frac{1}{Cn}(\phi^3 - \phi), \tag{15}$$

$$\frac{D\phi}{Dt^*} = \frac{1}{Pe_\phi} \nabla^2 \lambda^*, \tag{16}$$

$$\nabla \cdot (\sigma^* \nabla \varphi^*) = 0, \tag{17}$$

$$\rho^* C_p^* \frac{D\theta}{Dt^*} = \frac{k^*}{Ma} \nabla^2 \theta + J\sigma^* |\mathbf{E}^*|^2, \tag{18}$$

$$\nabla \cdot \mathbf{V}^* = 0, \tag{19}$$

$$\text{Re}\,\rho^* \frac{D\mathbf{V}^*}{Dt^*} = -\nabla p^* + \nabla \cdot \left[ \mu^* \left( \nabla \mathbf{V}^* + \nabla \mathbf{V}^{*T} \right) \right] + F_c \mathbf{F}_\mathbf{S}^* + \zeta \mathbf{F}_\mathbf{E}^*, \tag{20}$$

$$\mathbf{F}_\mathbf{E}^* = -\frac{1}{2}\left[ \left( \frac{\nabla \sigma^*}{\sigma^*} - \frac{\nabla \varepsilon^*}{\varepsilon^*} \right) \cdot \mathbf{E}^* \frac{\varepsilon^* \mathbf{E}^*}{1+\left(\omega^* \tau^*\right)^2} + \frac{1}{2}\left|\mathbf{E}^*\right|^2 \nabla \varepsilon^* \right]. \tag{21}$$

The nondimensionalization scheme gives some dimensionless parameters which are seen in the nondimensional form of equations, such as Cahn number: $Cn = \xi/H$; Peclet number: $Pe_\phi = u_0 H^2 / M_c \eta$; the value of $M_c \,(= Cl^4/\sqrt{\tilde{m}\varepsilon_e})$ is taken from Qian et al. [61] where $C\,(\sim 0.023)$ is a constant and $l$, $\varepsilon_e$, and $\tilde{m}$ are the length scale, energy scale, and molecular mass of the Lennard-Jones potential, respectively. Marangoni number: $Ma = \rho_0 C_{p0} u_0 H / k_0$; Reynolds number: $\text{Re} = \rho_0 u_0 H / \mu_0$; Thermocapillary force number: $F_c = \eta_0 / u_0 \mu_0$; Joule number $J = \sigma_0 \varphi^2 / k_0 \Delta T$; ACET force number $\zeta = \varepsilon_0 \varphi^2 / \mu_0 u_0 H$. The reference angular frequency is defined as: $\omega_0 = \sigma_0 / \varepsilon_0$. For electrothermal flow typical value of electrical conductivity is of the order of $10^{-2}$ S/m [45]. For an applied frequency $\omega_0 \geq 10^5$ Hz, the AC electroosmotic effects can be neglected [43,44,62]. Therefore, the dimensionless frequency is of the order of $\omega^* = 10^5 / (10^{-2}/10^{-9}) = 10^{-2}$. The dimensionless form of the charge relaxation time is: $\tau^* = \varepsilon^*/\sigma^*$.

All boundary conditions imposed on the boundaries to generate the fluid motion are shown in the Fig. 1(b). In order to consider the real part of the ACET forces, a phase difference of 180° is taken between adjacent electrode pairs [45]. To impose AC field in the bulk solution, voltages of $\varphi_{rms}^* = \pm 1$ are applied to the electrodes whereas other boundaries are electrically insulated. To obtained the thermal field, temperature of the inlet and outlet are considered as $\theta = 1$ and $\theta = 0$, respectively, while other boundaries are thermally insulated. For velocity field, standard no slip and no-penetration conditions are set on the walls. The inlet and outlet of the channel are set at zero gauge pressure since the present configuration of the binary system does not involve pressure-driven actuation. Initially, we consider zero velocity and zero temperature throughout the domain i.e., $V^* = \theta = \left(x^*, y^*, t^* = 0\right) = 0 \, \forall \, x^*, y^*$. From here onwards, we neglect superscript * to represent the parameters. All the values, data are in nondimensional form.

**C. Numerical methodology, model benchmarking and mesh independent study**

Temperature-driven moving contact line dynamics via electrothermal, thermocapillary, and surface tension forces is a multiphysics problem since convection-diffusion equations conjunction with electrical and energy equations occur simultaneously. Set of governing equations (Eqs. (15)-(21)) seems to be coupled and need to be solved simultaneously. COMSOL, a FEM based software packages and deals with multiphase-interface, is used to solve the multiphase flow dynamics. We have adopted Galerkin least square method to discretize the conservation equations. For time stepping generalized-$\alpha$ scheme and PARDISO solver are used in the simulations. We have set the tolerance levels of $10^{-6}$ to achieve high accuracy in the numerical results. An adequate finer meshing throughout the domain is adopted where mesh sizes of $\Delta x = \Delta y = 1.25 Cn$ are taken for the simulations. Before imposing any perturbation to the bulk solution an equilibrium and stable interface should be maintained. This can be achieved solving following equation:

$$\lambda(\phi) = 0. \tag{22}$$

The numerical framework is extensively benchmarked following dual benchmarking strategy. We have first benchmarked our numerical methodology with Liu et at. [58] whose investigations were based on Lattice Boltzmann phase field formalism. The numerical results are coupled with only thermocapillary actuated flow dynamics of two immiscible fluids. Using a imposed temperature gradient between top and bottom walls we show the temperature contours in Fig. 2(a) whereas Fig. 2(b) shows the contours, obtained results of Liu et al. [58]. An excellent agreement is achieved between our numerical solutions and with those of obtained in Ref. [58]. Next, we have benchmarked phase field framework of contact line dynamics with Wang et al. [29] whose numerical results were, in turn, benchmarked with MD simulations of Ref. [61]. Fig. 2(c) shows the data of contact line velocity over the alternative patched surface under pressure driven flow conditions of two immiscible fluids. It is evident that a good agreement is found between our obtained results and results of Ref. [29].

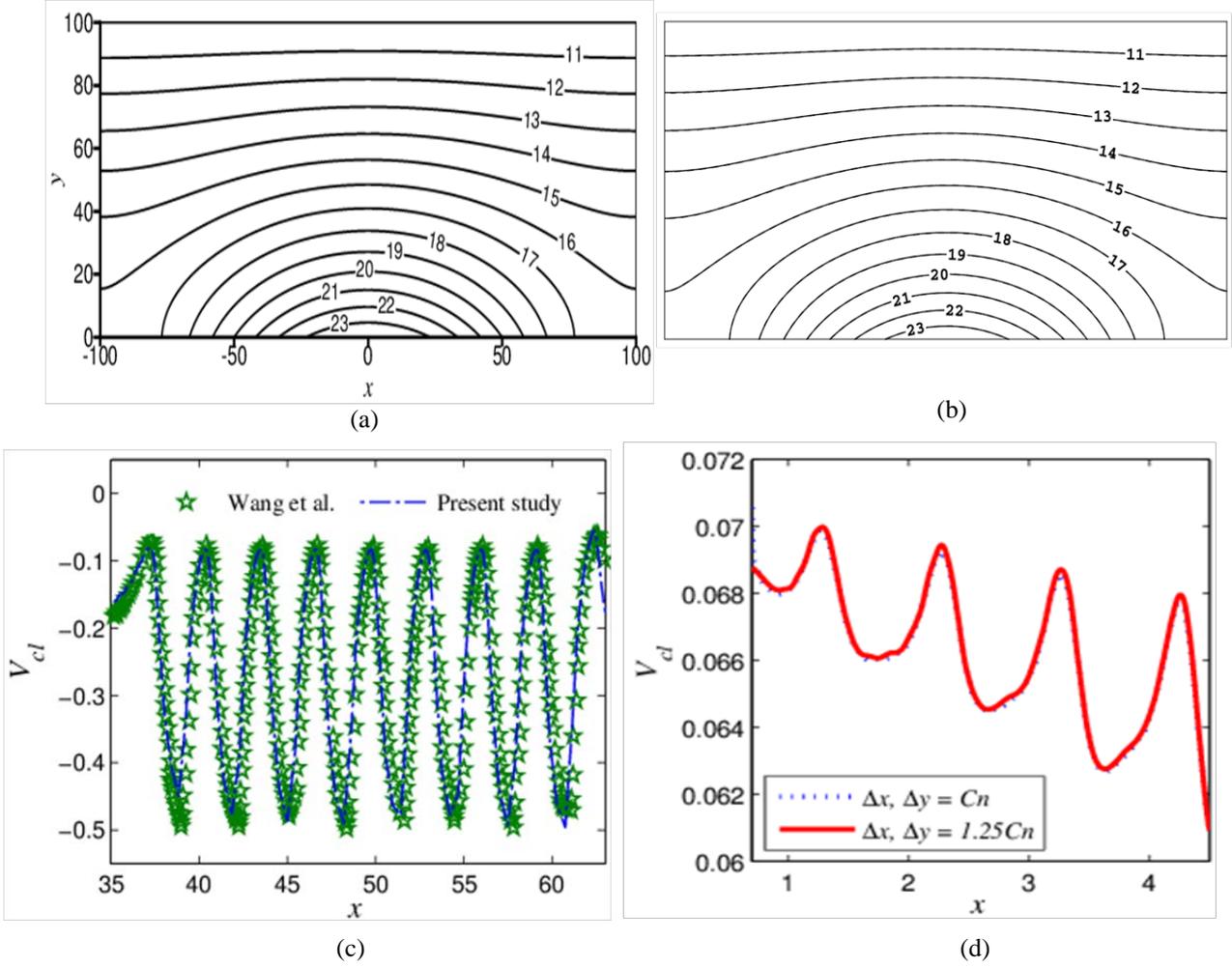

FIG. 2. Benchmarked solutions: temperature contours of two immiscible fluids (a) obtained from our study (b) reported in Liu et al. [58]. (c) Variation in contact line velocity with contact line location where data with star symbol show contact line velocity of Wang et al. [29] (reference), and the data with a dashed line are the results from our numerical framework. The results obtained using present numerical framework are well matched with references. (d) Mesh independent study: The contact line velocity with its position is highlighted for two different mesh sizes of $\Delta x, \Delta y = Cn$ and $1.25Cn$, the parameters used are $Cn = 0.01$, $Pe_\phi = 0.01$, $Ma = 0.01$, $\mathrm{Re} = 0.01$, $J = 0.01$, $B = \eta_T / \eta_0 = -0.01 \mathrm{K}^{-1}$, $F_c = 50, \zeta = 50$ and $\theta_s = 60°$. The deviation in contact line velocities for different mesh sizes is negligible. Therefore, the mesh of larger size is used for numerical solutions.

Phase field theory follows a sharp-interface approach which implies that the interface should be very thin, so that smooth transition of various forces across the interface can be taken. Typically for $Cn (= \xi / H) \to 0$, the phase field theory approaches the sharp-interface limit. However, in practice, below a threshold value, the distribution of the order parameter and other independent variables does not depend on the Cahn number [22,63]. This condition is also known as Cahn number independence test which directly portrays mesh independence test. We show the mesh independence study in the form of Cahn number independence test Contact line velocities for two different mesh (triangular) sizes (taken in the form of Cahn number) are presented in the Fig. 2(d). The various parameters involved in the results are shown in the figure caption. It is clear that deviation of the data of contact line velocity is negligible as the mesh size changes $\Delta x, \Delta y = Cn (= 0.01)$ to $\Delta x, \Delta y = 1.25 Cn$. Therefore, we have taken mesh sizes of $\Delta x, \Delta y = 1.25 Cn$ to simulate all the results in the present study.

## III. Results and discussions

In this section, we investigate the contact line dynamics and its implications on interfacial motion of the two phase system. As mentioned previously, our attention is on observation of alteration in capillary filling dynamics under temperature gradient-driven conditions where a number of forces interplay simultaneously. The thermocapillary forces and electrothermal forces are characterized by the thermocapillary force number and ACET force number, respectively, surface tension force is deeply altered with changing wettability condition which changes the local curvature of the interface. Therefore, we reveal the important characteristics of the flow dynamics on alteration of the following parameters: (1) thermocapillary force number ($F_c$), (2) ACET force number ($\zeta$) and (3) wettability of the surface i.e., static contact angle ($\theta_s$). The various dimensionless parameters in the transport equations are kept constant throughout the discussion, such as $Cn = 0.01, Pe_\phi = 0.01, Ma = 0.01, \text{Re} = 0.01, J = 0.01$ and $B = \eta_T / \eta_0 = -0.01 \text{K}^{-1}$. Other parameters are mentioned in the figure caption. The above values of the parameters are based on the properties of the base fluid i.e., fluid A which is considered as KCl electrolyte solution. Property ratios in the binary system are taken as unity. Various nondimensional length of the geometrical parameters are $L = 5$, $H = 1$ $g = d_1 = 0.1$, $d_2 = 0.4$, $2s = 0.4$ and $L_1 = 0.5$. A thorough review of contact line dynamics reveals that contact line dynamics and intricate behavior of the interfacial motion are best described by contact line velocity and capillary filling time. Accordingly, most of the results are described in the form of these parameters. Besides, the interface profile often carries important messages of the interface motion. We delineate some of the results by presenting contours of the interface.

As mentioned earlier, the contact line dynamics is controlled by altered thermocapillary forces and electrothermal forces. The gradients in surface tension, electrical conductivity and permittivity arise due to the externally applied temperature gradient in the capillary. At one side, surface tension gradient results in Marangoni stress across the contact line. On the other side, local variations in electrical conductivity and permittivity combined with electrical field generate electrical stresses across the contact line. Therefore, combined effects of these stresses set the motion into the binary fluids. The distributions of the order parameter, temperature, and potential lines along with gradients of surface tension, electrical conductivity and permittivity will determine the net forcing across the contact line.

To make an assessment of dominating characteristics of various forces we show contact line velocities ($V_{cl}$) with contact line position ($x$) for three different cases: (a) presence of only thermocapillary force, (b) presence of only electrothermal forces, and (c) thermocapillary forces and electrothermal forces are competitive in Fig. 3. In order to reveal distinctive characteristics of three cases, we consider same magnitudes of thermocapillary force number and ACET force number ($F_c = \zeta = 50$), other parameters used in the results are mentioned in the figure caption. It can be seen that contact line velocity for case (a) possesses a high value at the commencement of the motion, then it sharply decreases toward the downstream direction (Fig. 3(a)). Generally, interfacial tension of the fluids decreases linearly with temperature. Accordingly, a thermocapillary shear stress ($= d\eta/dx = (d\eta/dT)(dT/dx)$) is generated at the interface, which, in turn, triggers the fluids from a hotter region to colder region. Thermocapillary forces are directly influenced by the horizontal temperature gradient for a horizontal channel with hot and cold sides at two ends, vertical temperature gradient usually has no significant contribution on continuous motion of the fluids [64]. In our studies, a horizontal temperature gradient of $(T_h - T_c)/L$ is set across the channel imposing temperatures $T_h$ at inlet and $T_c$ at the outlet. Therefore, from entry region to end of the channel temperature gradient gradually decreases. The interface starts moving from a location of $L_1 = 0.5$ where the temperature gradient is sharp and interface/contact line velocity gets accelerated. However, during the course of the contact line motion, thermocapillary forces are decreased at areas of reduced temperature gradient. In contrast, electric field driven moving contact line motion does not show sharp decrement in contact line velocity. In Fig. 3(b) local contact line velocity and corresponding spatially averaged velocity ($V_{cl,avg}$, dotted line) are shown for case (b). Here, $V_{cl,avg}$ is defined as $V_{cl,avg}(l) = (1/(l - L_1)) \int_{L_1}^{l} V_{cl} \, dl$, where the distance $l = x - L_1$, At $x = L_1$, $V_{cl,avg}(L_1) = V_{cl}(L_1)$. The oscillatory electric field results in oscillation in contact line velocity. It is interesting that the amplitude of the oscillation increases as the contact line progresses towards the end of the channel. The gradually decreased thermal gradient in an AC field generates higher amplitude in the areas of reduced temperature gradient. It is quite notable that the position averaged velocity slowly decreases along the capillary. We mentioned previously that ACET driven flow is the consequence of the electric and thermal fields. In the present scenario, the temperature gradient reduces toward the outlet of the channel. However, non-uniform electric field prevailed into the full domain, does not allow sharp drop in the average velocity. This is interesting and unique to the ACET driven contact line dynamics. In addition, average contact line velocity with electric field is approximately three times of the contact line velocity in comparison with only thermocapillary forces. This is also advantageous for electrothermal driven binary fluids compared to the thermocapillary driven binary fluids. When we superimpose the two effects on the contact line motion it can be investigated that still flow oscillations are prevailed in the domain, hence, undulation in contact line velocity is observed (Fig. 3(c)). This is obvious because for same values of two force numbers i.e., thermocapillary force number and ACET force number, ACET effects are more dominant compared to thermocapillary effects as we have seen before. Therefore, electric field driven contact line motion shows dominant effects over thermocapillary driven contact line motion with the considerations that all the boundary conditions and force amplitudes are identical. As leading characteristics of thermocapillary effects which can be revealed at this juncture that oscillatory local velocity and its spatially averaged velocity sharply fall during the traveling of the contact line along downstream location. This sharp decreasing in contact line velocity is one of the dominating characteristics of thermocapillary effects as seen in Fig. 3(a). The contact line velocity for case (c) where both thermocapillary and electrothermal motions are adopted shows higher velocity in comparison with cases where individual effects are considered. Therefore, presence of a thermal field in an AC frequency domain results in thermocapillary and electrothermal actuations which aid with each other.

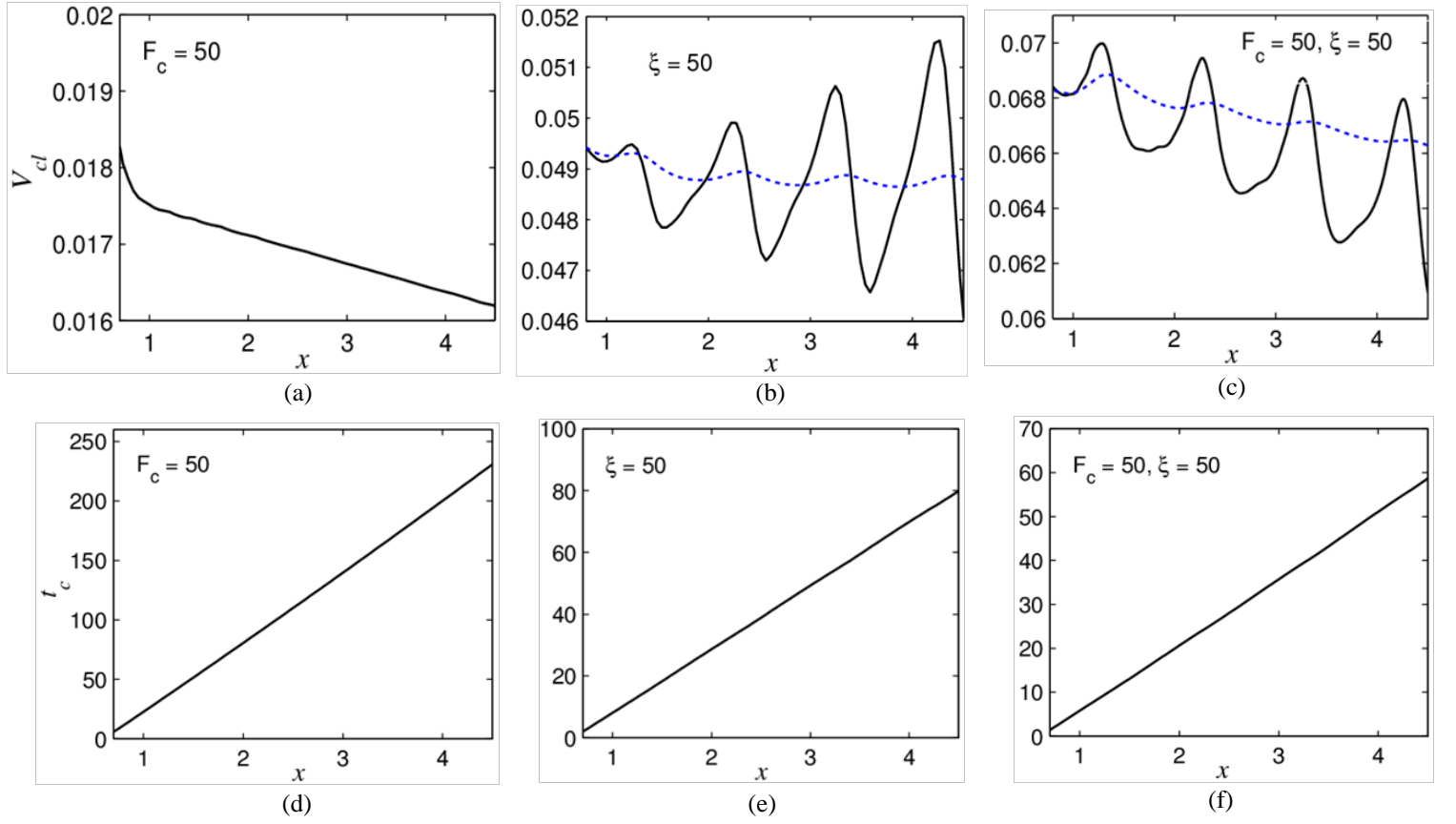

FIG. 3. (a)-(c) variation in contact line velocity ($V_{cl}$) with contact line location (x); (d)-(e) plots of capillary filling time ($t_c$) with position. $F_c = 50, \zeta = 0$ for (a) and (d); $F_c = 0, \zeta = 50$ for (b) and (e); $F_c = 50, \zeta = 50$ for (c) and (f). For all cases, surface wettability is $\theta_s = 60°$. For equal magnitude of force numbers, electrothermal effects are more prominent for high contact line velocity. Accordingly, capillary filling time also long for the case of thermocapillary flow. The rate of decrement in contact line velocity is much sensitive for thermocapillary flow. This effect also influences the contact line velocity for the case of combined activation of thermocapillary and electrothermal forces.

The important distinctive features drawn in the previous paragraph can be assessed precisely on capillary filling time ($t_c$) vs position plots (Figs. 3(d)-(f)). It is evident that the time to traverse the channel is much longer ($t_c \approx 225$) for case (a) when only thermocapillary forces actuate the two phase flow. However, filling time drops appreciably with electrothermal forces. As discussed electrothermal forces result in almost three times contact line velocity than the thermocapillary actuation. The filling time is also reduced by a factor of 3 compared to thermocapillary driven flow. For the cases of dominating thermocapillary and electrothermal forces, one can observe the shortest time ($t_c \approx 58$) to travel the channel.

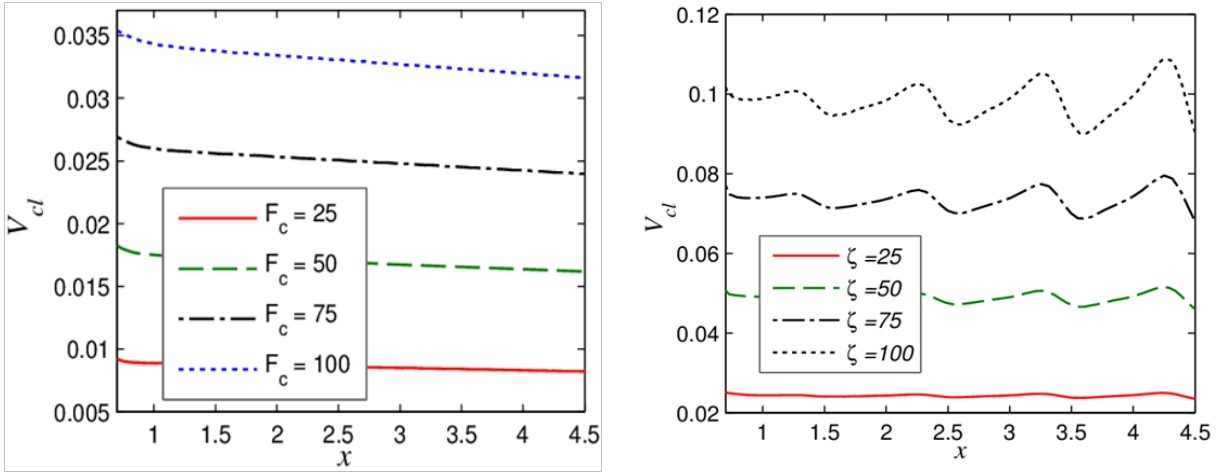

FIG. 4. Plots of contact line velocities with its position for (a) $F_c = 25, 50, 75$ and $100$, other parameters are $\zeta = 0$ and $\theta_s = 60°$. (b) $\zeta = 25, 50, 75$ and $100$, other parameters are $F_c = 0$ and $\theta_s = 60°$. With increasing thermocapillary number contact line velocity increases. However, at equal numbers, contact line velocity for electrothermal effects is greater than the contact line velocity for thermocapillary effects. The oscillatory electric field results undulations in fluid motion. Therefore, contact line velocities are adjusted with undulations.

To explore the competition between two forces in a more involved way we represent the contact line velocity plots with its position for different thermocapillary force number ($F_c$) and ACET force number ($\zeta$) separately in Fig. 4. The various parameters used in the results are shown in the figure caption. It is evident that for all values of $\zeta$ and corresponding $F_c$, the electrothermal mechanism is more effective in enhancing the contact line velocity. Also, average contact line velocities for the electrothermal process are almost constant whereas for thermocapillary flow it is gradually decreasing. The trend of variations of the contact line velocities shows an incremental effect on velocities with increasing force numbers. This is obvious because both the numbers are directly multiplied with body force expression (see Eq. (20)). Therefore, increasing $F_c$ and $\zeta$, effectively the driving force to drag the contact line over the substrate is increased which results in strong boosting on the interface to move along the channel. The assessment of the influence of these force numbers can also be inferred following a different way. Thermocapillary force number is the relative strength of thermocapillary forces over viscous force. Therefore, thermocapillary stress across the interface increases to speed up the contact line velocity with increasing thermocapillary number. On the other hand, ACET force number comprises a square of applied voltage in the numerator. Keeping fluid properties as constant increasing ACET force number signifies increase in imposed voltages on the electrode pairs. At higher voltage higher electric field strength results in stronger actuating forces in the bulk solution. Hence, the interface/contact line moves at a faster speed with increasing ACET force number.

In the previous paragraphs, we have investigated the interaction between two dominating forces, namely thermocapillary and electrothermal forces in the context of contact line dynamics in presence of imposed thermal field in the channel. For all the results we consider predefined wettability of static contact angle $\theta_s = 60°$ on the surfaces of the channel. Previous investigations of Wang et al. [29] on moving contact line dynamics of two immiscible fluids over alternative two different patches revealed that alteration in wettability periodically changes the surface energy in an intriguing manner via diffusion which results in facilitating variation in contact line velocity. As per requirement, contact line velocity can be accelerated or decelerated through stick-slip motion. Keeping this, in mind, we investigate the influence of wetting characteristics of the surfaces on interfacial dynamics of the two phase system. Fig. 5 shows the effect of wettability condition on the variation of contact line velocity and capillary filling time with contact line position for different wettabilities $\theta_s = 60°$, 80°, 100°, 120° and 140°. Other parameters involved in the analysis are mentioned in the figure caption. From the figure, one can see a significant increment in contact line velocity with decreasing contact angle. The surface has strong affinity to attract the displaced fluid (fluid B) for $\theta_s > 90°$. As the obtuse contact angle differs from 90° the affinity becomes stronger to stick the fluid B and induces a resistance on the progression of the contact line. The altered surface energy causes net interfacial forces which opposes the driving electrothermal forces. On the other hand, forcing direction becomes reverse for $\theta_s < 90°$. In this scenario, wetting characteristics of the surfaces repel the displaced fluid B and attract the displacing fluid A. Therefore, modulated surface energy introduces net interfacial forces acting along the downstream direction along which the ACET forces act. As the acute contact angle differs from 90° forcing parameter becomes more effective to push the binary fluids towards the end of the channel. Therefore, the ranges of static contact angle in which the surface wettability facilitate or inhibit the movement of the interface along the forward direction are $\theta_s < 90°$ and $\theta_s > 90°$, respectively, and positively acting interfacial forces decreases as $\theta_s$ approaches 90°. A typical example of oppositely acting thermocapillary forces is exhibited as follows. After a sufficiently long time ($t_c = 200$) contact line traverses a very short distance of $x \approx 1.7$ for $\theta_s = 140°$. For this case, the contact line speed is very low $V_{cl} \approx 0.01$. In contrast, contact line achieves the highest velocity of $V_{cl} \approx 0.07$ at $\theta_s = 60°$

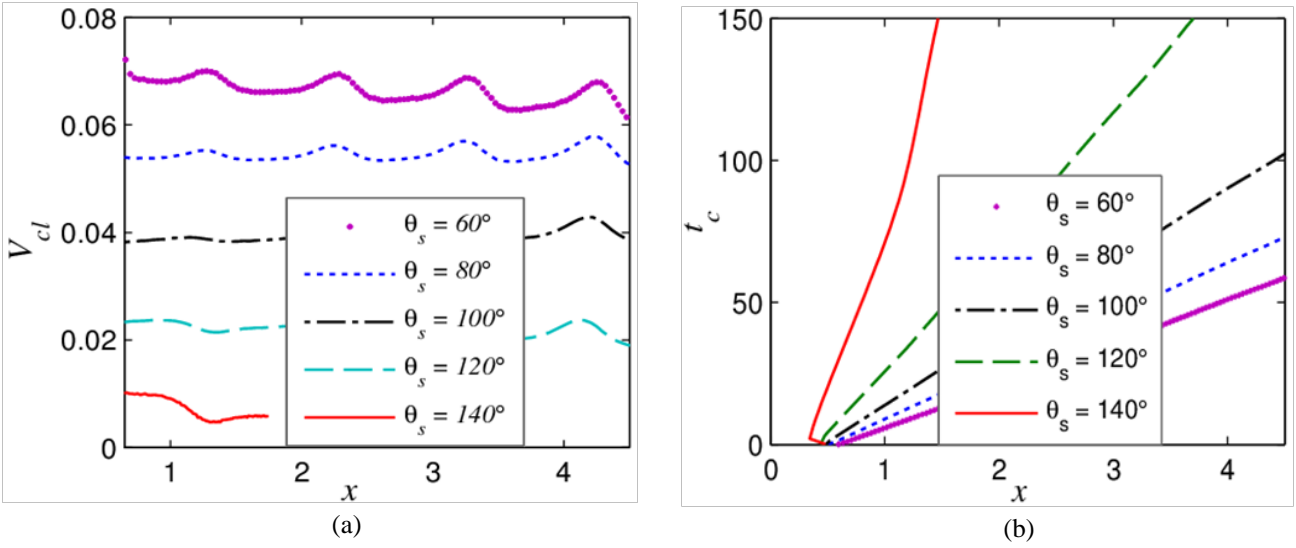

(a)  (b)

FIG. 5. Plots of (a) contact line velocities and (b) capillary filling times for various wetting conditions: $\theta_s = 60°$, 80°, 100°, 120°, and 140°. Other parameters used in the results are $F_c = 50$ and $\zeta = 50$. The interaction between surface and fluids becomes important on altered static contact angle. The local curvature effects change the surface tension forces, and hence, net interfacial forces. With increasing static contact angle the surface tension force which aids the driving forces decreases till $\theta_s = 90°$. Then, it starts opposing on increasing $\theta_s$. In this range, the net interfacial force opposes the contact line motion to move toward the end of the channel.

All the important characteristics which were mentioned before can be assessed in the filling time vs position plot (Fig. 5(b)). One can see filling time appreciably increases with increasing static contact angle. The chemical properties of the surface alter the surface energy which causes an alteration in surface tension force across the contact line. At lower contact angle net interfacial force aid the driving electrothermal forces and contact line traverse the channel at a faster rate. On the other hand, at larger contact angle net interfacial force change its direction and starts opposing the ACET forces. Therefore, the speed of the contact line reduces and the time to travel the channel is increased. For $\theta_s = 60°$ the contact line traverses the entire length of the channel at $t_c \approx 60$, while for $\theta_s = 100°$, the traverse time increases to tc = 100. for a further increase in contact angle to $\theta_s = 120°$ results in severe flow retardation and the contact line traverses only till $x \approx 1.5$ at $t_c = 150$.

We have seen above a threshold static contact angle ($\theta_s > 90°$) the net interfacial force changes its direction and pull the bulk fluid toward the entry of the channel. It was also found that identical thermocapillary force number and ACET force number do not show equal strength to generate same contact line velocities. For a given thermocapillary force with predefined contact angle, there is a corresponding ACET force number which will exhibit equal strength for transporting the binary fluids. To figure out the overall balance between these forces we display the contact line velocity and capillary filling time plots for various ACET force number and surface wetting conditions in the range of $\theta_s > 90°$ where interfacial tension forces oppose the ACET forces and create a reverse movement of the interface (Fig. 6). For all cases, we consider thermocapillary force number as $F_c = 50$ the relevant other parameters are shown in the figure caption. An important point to be mentioned that for clear visualization of backward-forward motion we take the initial position of the interface as $x = 2.5$, the half of the channel length. In Fig. 6(a) and 6(d), the results are shown for wettability of $\theta_s = 100°$ where resisting behavior of the interfacial tension forces against the electrothermal forces is not strong. For ACET force number $\zeta = 30$ and 50 one can see interface travels full-length of the channel. However, for $\zeta = 10$ the oppositely acting forces are almost balanced and a very low velocity along the downstream direction prevails. Corresponding time vs position plot also depicts a clear scenario of these events. Dominating ACET forces over the interfacial forces impose adequate impact across the contact line and push toward the entry of the channel for the cases of $\zeta = 30$ and 50. However, after a sufficiently long time contact line traverses a very small distance of $\Delta x \approx 0.25$ for $\zeta = 10$. Therefore, in our investigations opposing interfacial forces for the thermocapillary number $F_c = 50$ and surface wettability of $\theta_s = 100°$ are almost balanced by ACET forces having ACET force number $\zeta = 10$. Similar observations on contact line velocity and filling time plots for static contact angle $\theta_s = 120°$ reveal that equally dominating characteristics of the various forces generated from the thermal field are shifted to a higher value of ACET force number. Previously, it was observed that the net interfacial forces which are activated along backward direction become stronger at higher static contact angle. Accordingly, the thermocapillary force of $F_c = 50$ is almost balanced with electrothermal forces with $\zeta = 30$. Above and below this critical limit of $\zeta$, electrothermal forces dominate over the interfacial forces and interfacial forces dominate over electrothermal forces, respectively. As a consequence, one can see forward and backward motions of the contact line along the channel. The relevant filing time vs axial position plot (Fig. 6(e)) also reveals those important dynamical characteristics of the contact line. Typically, for an equilibrium configuration of various forces contact line moves a very short distance of $\Delta x \approx 0.5$ during the course of contact line motion for a time span $\Delta t = 200$. A closer look on Figs. 6(a) and 6(b) shows that the magnitude of contact line velocity for $\theta_s = 120°$ is approximately half of the velocity of $\theta_s = 100°$, for both cases $\zeta = 50$. With increase in surface affinity from $\theta_s = 100°$ to $\theta_s = 120°$ the oppositely acting increased interfacial tension forces result in lower contact line velocity. A different scenario is investigated when the static contact angle is further increased to $\theta_s = 140°$. Here, net interfacial force dominates over ACET forces for $\zeta = 10$ and 30 and for both cases one can see the backward movement of the contact line. However, interfacial forces are equilibrated with its resistive force at $\zeta = 50$, for which, the traveling distance is negligible. Corresponding capillary filling time with contact line position plots precisely shows the important characteristics for $\theta_s = 140°$. At moderate time duration ($t \approx 75$) contact line reaches to the inlet of the channel following backward traversing for $\zeta = 10$. In contrast, at higher ACET force number ($\zeta = 50$) the balanced forcing parameters do not allow motion along either of the directions. Having a closer look at Figs. 6(b) and 6(c) we see an increase in the interfacial tension force on alteration of wetting characteristics from $\theta_s = 120°$ to $\theta_s = 140°$, which results in doubling the velocity which causes backward motion of the interface. An important point to be noted that when we initiate the various forces electrothermal forces take some time to activate whereas interfacial forces are high at the initial stage and imposed sudden perturbation into the simulation domain, especially across the fluid-fluid-solid contact line where diffusion is much higher. This phenomenon at initial stage cause abrupt increases in contact line velocity. However, interfacial forces are balanced on activation of electrothermal forces after the short time duration and smooth progression of the interface through the channel is observed.

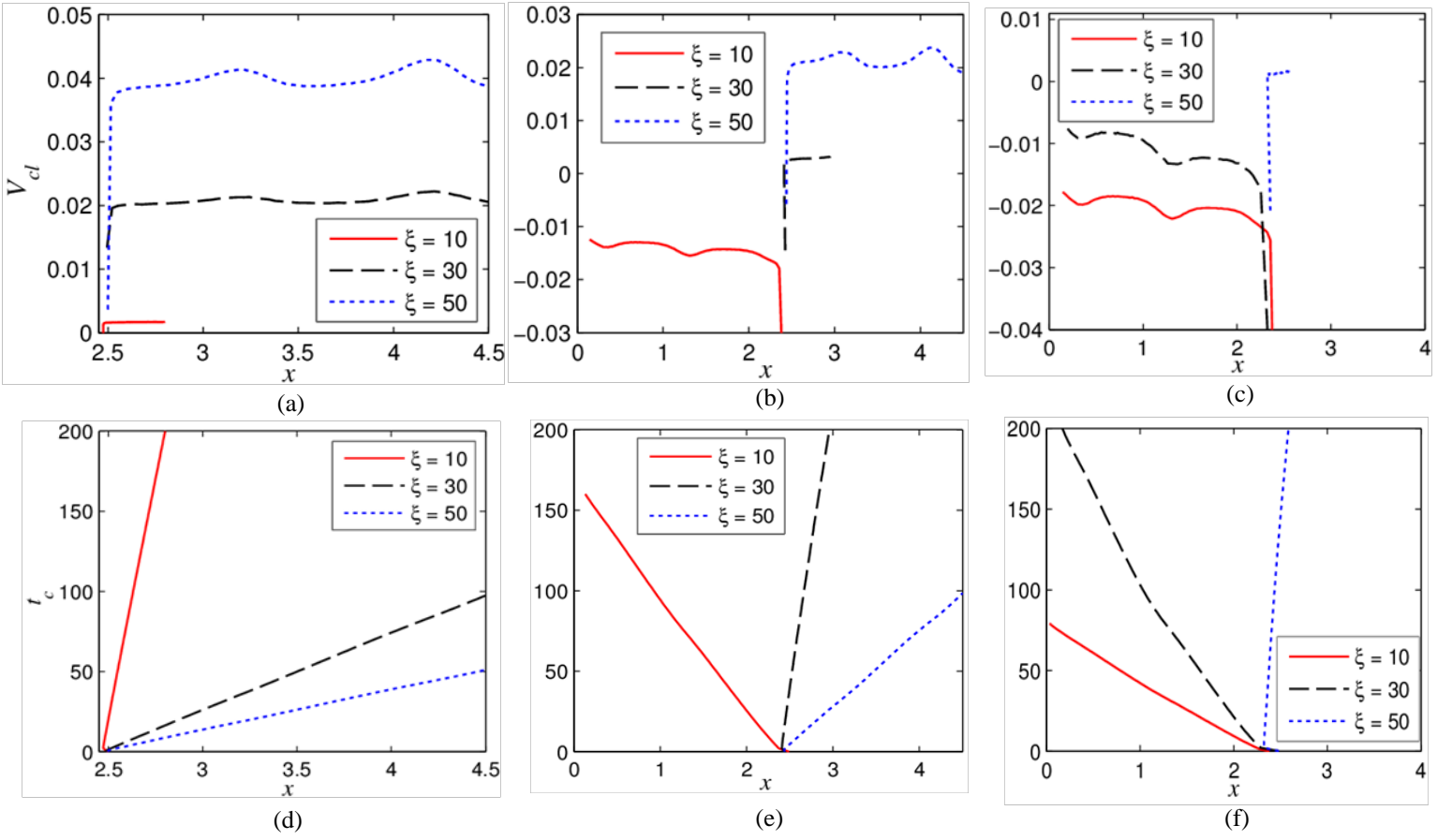

FIG. 6. (a)-(c) variation in contact line velocity ($V_{cl}$) with contact line location (x); (d)-(e) plots of capillary filling time ($t_c$) with position. $\theta_s = 100°$ for (a) and (d); $\theta_s = 120°$ for (b) and (e); $\theta_s = 140°$ for (c) and (f). For all cases, thermocapillary force number is $F_c = 50$. Relative dominating characteristics of interfacial tension force with other forces is clearly visualized. For fixed thermocapillary force number, the interfacial tension force is dominating at low ACET force number. However, At sufficiently high ACET force number electrothermal force dominates and the contact line moves toward the end of the channel. The strength of the interfacial tension force is increased with increased imposed static contact angle. All the important characteristics of the velocity plots are explicitly attributed in time vs position plots.

## IV. Conclusion

We show our numerical results of interfacial dynamics of contact line motion of a binary system constituting two immiscible fluids. The flow is actuated by the combined effects of thermocapillary and alternating current electrothermal forces in a narrow channel whose surfaces are chemically treated with predefined wettability gradient. An externally applied temperature gradient is applied across the channel to raise the gradients in surface tension, permittivity and electrical conductivity which are responsible to generate stresses across the interface and to evolve contact line motion. Our primary focus is to control the contact line motion and its spatiotemporal dynamics through controlling the various forces which directly alter the net force across the contact line. Accordingly, we defined two important nondimensional parameters: thermocapillary force number and ACET force number which depict the relative strength of thermocapillary and electrothermal forces. Through a parametric study, we showed that contact line motion and its intricate flow physics can be effectively tuned on altering the governing parameters associated with present flow configuration. For same thermocapillary force number and ACET force number, ACET forces cause higher contact line velocity compared to thermocapillary forces. The wetting characteristics alter the direction of the interfacial forces, which, in turn, results in backward motion at lower ACET force number. The investigations inferred in the study have the potential to build the fundamental understanding of a multiphase flow dynamics involving thermal actuation of the binary system.